# Strong evidence of low levels of solar activity during the Maunder Minimum

V.M.S. Carrasco[1,2,*] • H. Hayakawa[3,4,5,6] • C. Kuroyanagi[7] • M.C. Gallego[1,2] • J.M. Vaquero[2,8]

[1] Departamento de Física, Universidad de Extremadura, 06006 Badajoz, Spain.

[2] Instituto Universitario de Investigación del Agua, Cambio Climático y Sostenibilidad (IACYS), Universidad de Extremadura, 06006 Badajoz, Spain.

[3] Institute for Advanced Researches, Nagoya University, Nagoya, 4648601, Japan.

[4] Institute for Space-Earth Environmental Research, Nagoya University, Nagoya, 4648601, Japan.

[5] UK Solar System Data Centre, Space Physics and Operations Division, RAL Space, Science and Technology Facilities Council, Rutherford Appleton Laboratory, Harwell Oxford, Didcot, Oxfordshire, OX11 0QX, UK.

[6] Nishina Centre, Riken, Wako, 3510198, Japan.

[7] Graduate School of Arts and Sciences, University of Tokyo, Tokyo, 1538902, Japan.

[8] Departamento de Física, Universidad de Extremadura, 06800 Mérida, Spain.

* Corresponding author: V.M.S. Carrasco (vmscarrasco@unex.es)

**Abstract:** The Maunder Minimum (MM) was a period of prolonged solar activity minimum between 1645 and 1715. Several works have identified a significant number of problematic spotless days in the MM included in existing databases. We have found a list of exact spotless (in the second half of 1709) and spot days (January and August 1709) provided by Johann Heinrich Müller. We computed the most probable value and upper/lower limits of the active day fraction (ADF) from Müller's data using the hypergeometrical probability distribution. Our sample is not strictly random because Müller recorded observations in consecutive days when he observed sunspots. Therefore, our result represents an upper threshold of solar activity for 1709. We compared this result with annual values of the ADF calculated for the Dalton Minimum and the most recent solar cycles. We concluded that it was less active than most years both in the Dalton Minimum and in the most recent solar cycles. Therefore, the solar activity level estimated in this work for 1709 represents robust evidence of low solar activity levels in the MM.

## 1. Introduction

A prolonged solar activity minimum occurred in the years of 1645–1715 (Soon & Yaskell 2003; Arlt & Vaquero 2020). After examining solar activity records such as sunspots, auroras and cosmogenic isotope data, Eddy (1976) concluded that this grand minimum period was a real feature of the history of the Sun and not a limitation in terms of observational capabilities. He named it the Maunder Minimum (MM), appreciating the scientific achievements of Edward Maunder on this prolonged solar activity minimum (Maunder 1894, 1922). More recently, some studies have questioned the level of solar activity during the MM (Zolotova & Ponyavin 2015), but these have been subjected to immediate criticisms and rejections (Usoskin et al. 2015). Thus far, the MM is considered the only grand minimum period within the telescopic era, in contrast with the Dalton Minimum (Hayakawa et al. 2020a, 2021a). Accordingly, its study is fundamental to understanding long-term solar activity and the solar–terrestrial relationship (Vaquero & Vázquez 2009; Owens et al. 2017; Muñoz-Jaramillo & Vaquero 2019).

Hoyt & Schatten (1998) compiled numerous sunspot records covering also the MM to build the Group Sunspot Number index. They obtained an observational coverage greater than 95% for that period. However, more recent studies found significant numbers of problematic sunspot records in the database of Hoyt & Schatten (1998) due to contaminations from astrometric observations and general descriptions (e.g., Clette et al. 2014; Vaquero & Gallego 2014; Carrasco, Villalba Álvarez & Vaquero 2015; Usoskin et al. 2015; Hayakawa et al. 2021b). Vaquero et al. (2016) removed some of these problematic observations detected in previous works for the revised collection of sunspot group numbers. In addition to the absence of sunspots (Hoyt & Schatten 1998; Vaquero et al. 2016), this period presents other particular characteristics such as asymmetric sunspot occurrences in the southern solar hemisphere (Ribes & Nesme-Ribes 1993; Vaquero, Nogales & Sánchez-Bajo 2015a) and apparent loss of solar coronal streamers (Hayakawa et al. 2021a). This is contrasted with the nearly symmetric sunspot occurrences just prior to the MM (1642–1645) recorded in Hevelius' sunspot drawings (Carrasco et al. 2019).

Another outstanding fact is the MM's anomalous cycle length. No signal of the 11-year solar cycle during the MM appears in the Group Sunspot Number by Hoyt & Schatten (1998). Instead, a nine-year solar cycle was found by Vaquero et al. (2015b) using a subset of the Hoyt & Schatten (1998) database and Usoskin, Mursula & Kovaltsov (2000) determined a 22-year periodicity for sunspots during the MM. Recent analyses of cosmogenic isotope data have shown similar cyclicity (Usoskin et al. 2021), whereas one of the previous results obtained a 14-year solar cycle during the MM (Miyahara et al. 2004). This exceptionally lengthy cyclicity was claimed probably because one or two solar cycles could have been lost at the beginning and end of the MM from these cosmogenic isotope data (Owens, Usoskin & Lockwood 2012; Vaquero et al. 2015b).

Still, Carrasco et al. (2018) found that the umbra-penumbra ratios obtained from sunspots recorded in approximately 200 drawings during the MM are comparable to values calculated from modern sunspots. Accordingly, the absence of sunspots during the MM cannot be explained by changes in the umbra-penumbra ratios.

Within the telescopic era, other periods of reduced solar activity have been recorded (Silverman and Hayakawa 2021). The Dalton Minimum was a period that occurred around the first third of the 19th century with small solar cycles in terms of amplitude. However, although solar activity was low during the Dalton Minimum, this period is not considered to be a grand minimum period such as the MM because of its clearer cycle amplitudes (Sokoloff 2004; Vaquero et al. 2016; Hayakawa et al. 2020b). Moreover, sunspots occurred in both solar hemispheres during the Dalton Minimum (Hayakawa et al. 2020b, 2021c), although they mainly appeared in the northern hemisphere just before this period (Usoskin et al. 2009). The solar coronal streamers were visible (Hayakawa et al. 2020a), in contrast to what was observed during the MM (Hayakawa et al. 2021b). Recently, several studies, such as those of Fisher's, Hallaschka's, Derfflinger's, and Prantner's sunspot observations, have provided new information on this period (Denig & McVaugh 2017; Carrasco et al. 2018b; Hayakawa et al. 2020b, 2021c). Other outstanding periods regarding low solar activity are Solar Cycle 14 (1902–1913) and the modern Solar Cycle 24 (2009–2019). According to 13-month smoothed monthly data from the international sunspot number index (version 2) provided by the Sunspot Index and Long-term Solar Observations (SILSO; http://www.sidc.be/silso/; see Clette & Lefèvre, 2016), Solar Cycle 14 (107.1) and 24 (116.4) are the two solar cycles with the smallest maximum amplitudes since the Dalton Minimum. In addition, the amplitude corresponding to minima of Solar Cycle 24 and the current Solar Cycle 25 are the lowest amplitudes according to the sunspot number index (version 2) from the Dalton Minimum.

In this study, we use Johann Heinrich Müller's sunspot observations from 1709, which have been recently acquired and exploited by the Eimmart Collection of the National Library of Russia (Hayakawa et al. 2021d). On this basis, we calculate the solar activity level in 1709 and compare it with that during the Dalton Minimum and the most recent solar cycles. The remainder of this paper is organized as follows. In Section 2, we analyse the solar activity level in each year of the MM according to the number of active days and active day fraction (ADF). In Section 3, we present Müller's sunspot observation from 1709 and calculate the most probable value and its limits for the ADF from Müller's records using the hypergeometrical probability distribution. Section 4 compares results obtained from Müller's records corresponding to 1709 and the solar activity level recorded during the Dalton Minimum and the most recent solar cycles. We present our main conclusions in Section 5.

## 2. Solar Activity in 1709 and the Maunder Minimum

The year 1709 corresponds to the declining phase of the last solar cycle in the MM. Figure 1 (top panel) shows the daily numbers of sunspot groups recorded by all observers included in the existing sunspot group number database (Vaquero et al. 2016) for the period 1645–1715. Sunspot records dated in 1709 are shown in red. The total number of active days recorded in 1709 was 63, whereas the number of quiet days was 236 according to Vaquero et al. (2016). We consider an active day to be one when at least one sunspot was observed by any observer and a quiet day when all observers recorded 'zero' sunspots. Thus, the ADF for 1709 was 21.1%. Figure 1 (bottom panel) represents the number of active days (black dots) and the ADF (blue line) for the period 1645–1715 according to Vaquero et al. (2016). Regarding the number of active days, 1709 was the ninth most active year of the MM. The year 1684 was the only year of the MM belonging to the 17th century with a number of active days above that in 1709. The most active years according to the annual number of active days were 1705, 1707 and 1704 with 143, 115 and 106 active days recorded, respectively. According to the ADF, 1709 was the 13th most active year of the Maunder Minimum. However, we note that the three most active years in this case were years with very few observations and with practically no quiet days recorded: (i) 1645 with only one record and an ADF of 100%, (ii) 1656 with 15 records and an ADF of 86.7%, and (iii) 1655 with 26 records and an ADF of 50%. Apart from these three years, 1705, 1707 and 1676 were the years with the highest ADF values: 46.4%, 39.1% and 37.6%, respectively. Moreover, no sunspot records were produced in five years of the early MM (1646, 1647, 1649, 1650 and 1651). From these facts, we can conclude that 1709 was one of the most active years in the MM.

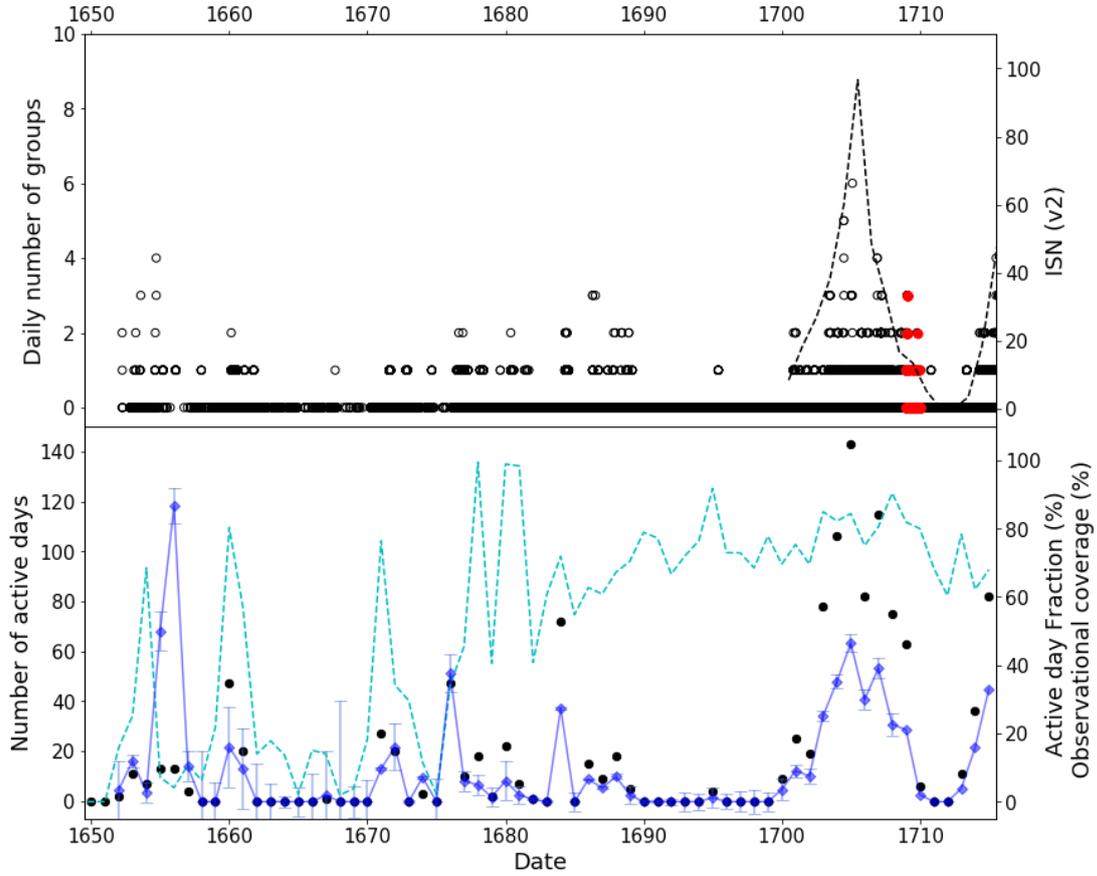

**Figure 1.** (Top panel) Daily number of sunspot groups recorded during the MM (1650–1715) according to Vaquero et al. (2016), additionally including corrections made by Hayakawa et al. (2021d) from sunspot observations made at the Eimmart Observatory are represented by black open circles (sunspot records for 1709 in red filled circles). The annual values of the international sunspot number index (version 2) from 1700 are depicted by dashed line. (Bottom panel) Number of active days (black dots) and ADF (blue diamond line) for the period 1650–1715 according to Vaquero et al. (2016) additionally including corrections made by Hayakawa et al. (2021d) from sunspot observations made at the Eimmart Observatory. Error bars were calculated from the hypergeometrical probability distribution. The cyan dashed line represents the annual observational coverage.

## 3. ADF in 1709 from Müller's sunspot observations

We found a list of spotless days recorded by J. H. Müller in his logbooks for 1709. In this list, 24 quiet days were recorded, all of them in the second half of 1709 (starting from 30 August). In addition, Müller recorded eight active days (5–9 January and 24, 25 and 27 August). Thus, the number of observation days in Müller's records for 1709 was 32. Regarding the period of Müller's observations in 1709 (5–9 January and 24 August – 30 December), the observational coverage was 17.4% when considering all days in those six months of January and from August to December. We acknowledge that this represents a low observational coverage (approximately 10% if we consider all observations for the entirety of 1709), whereas these observations were temporally well distributed. Regarding the sunspot observations made from 24 August to 30 December, the average between two consecutive observations was 4.96 days, and the maximum number of days between two consecutive observations was 20 days (from 28 October to 17 November). In addition, as this dataset showed both active and quiet days, we could calculate the most probable value of the ADF for 1709 together with the upper and lower limits from the samples by Müller of $n$ observations (32) with $r$ active days (8) using the hypergeometrical probability distribution (Kovaltsov, Usoskin and Mursula 2004):

$$p(s) = \frac{s!\,(N-s)!}{(s-r)!\,(N-s-n+r)!} \frac{n!\,(N-n)!}{(n-r)!\,N!\,r!}$$

where $N$ is the number of days in a year (365) and $s$ is the total number of active days within the year which is to be estimated. Thus, considering a 99% significance level, we determined that the most probable value for the ADF in 1709 was 25.1%, and the upper and lower limits were 38.9% and 11.3%, respectively. We should note that our sample was not strictly random because a greater number of consecutive observation days were made when Müller recorded only a single sunspot. For example, Müller observed every day from 5 to 9 January when he observed a single sunspot. Therefore, these values represent an upper limit of solar activity.

**4. Comparison of levels of solar activity in 1709 and other periods**

We compared the solar activity levels calculated from J.H. Müller's records in 1709 during the MM with those of the Dalton Minimum and the most recent solar cycles. Regarding the Dalton Minimum, we computed the ADF for each year of the period 1798–1823 according to data included in Vaquero et al. (2016) with revisions by Hayakawa et al. (2020b) (green triangles in Figure 2, top panel). The horizontal black lines indicate the most probable value along with upper and lower limits, corresponding to the 99% confidence interval, of the ADF calculated according to Müller's sunspot records from 1709. The most probable value for 1709 would be in the 20th percentile in a ranking of the most active years with respect to the Dalton Minimum. Thus, only six years in the Dalton Minimum (1798, 1809, 1810, 1811, 1822 and 1823) were less active than the most probable value of the ADF for 1709. The upper and lower limits of the ADF for 1709 would be around the 40th and 10th percentile (i.e., only four years were less active than the lower limit of the ADF for 1709, namely, 1809, 1810, 1811 and 1823, and eleven years in the case of the upper limit). Thus, 1709 was less active than most years in the Dalton Minimum.

We computed the ADF for years of the most recent solar cycles (1996–2019) to compare with that calculated from Müller data. We must apply some constraints to modern sunspot records to level to Müller's observations due to possible limitations in the observations made by Müller (for example, a smaller telescope would detect less sunspots). Thus, we use the sunspot catalogue published by the Kislovodsk Mountain Astronomical Station of the Central Astronomical Observatory at Pulkovo (Otkidychev & Skorbezh 2014; Muñoz-Jaramillo et al. 2015; Tlatov et al. 2019; Mandal et al. 2020) (http://158.250.29.123:8000/web/Soln_Dann/) to carry out this comparison. The Kislovodsk dataset is used as one of the reference datasets for the SILSO sunspot number (Mathieu et al. 2019). The ADF values obtained from raw Kislovodsk data are similar to those calculated from the SILSO data for the period 1996–2019. Table 1 shows these calculations. However, we note a minor difference in the results. Whereas the solar minimum in 2008 (24.3%) would be slightly deeper than that in 2019 (26.8%) from the raw Kislovodsk data, this is in contrast to the SILSO data, where the ADF calculated was 24.9% for 2019 and 27.6% for 2008. This difference is not statistically significant since, using the F-test, the F calculated value (0.02) is lower than F table value (4.05) for $p = 0.05$. We note that 16 quiet days (four active days) recorded at the Kislovodsk Observatory in 2008 and four quiet days (nine active days) in 2019 were active (quiet) days according to the SILSO data. This was the main reason for the difference between the ADF values for those years. Then, according to raw ADF values, only 2008, 2009 and 2019 would be comparable to the most probable value of the ADF for 1709, and 2018 would be comparable to its upper limit. No level of solar activity occurring in the modern years would be comparable to the lower limit of the ADF for 1709.

**Table 1.** Active day fraction (%) for the period 1996–2019 obtained from Kislovodsk (KADF) and SILSO (SADF) data.

| YEAR | KADF (%) | SADF (%) | YEAR | KADF (%) | SADF (%) |
|---|---|---|---|---|---|
| **1996** | 51.0 | 54.9 | **2008** | 24.3 | 27.6 |
| **1997** | 82.2 | 83.3 | **2009** | 26.7 | 28.2 |
| **1998** | 100.0 | 99.2 | **2010** | 86.4 | 87.9 |
| **1999** | 100.0 | 100.0 | **2011** | 97.7 | 99.5 |
| **2000** | 100.0 | 100.0 | **2012** | 99.4 | 100.0 |
| **2001** | 100.0 | 100.0 | **2013** | 100.0 | 100.0 |
| **2002** | 100.0 | 100.0 | **2014** | 100.0 | 99.7 |
| **2003** | 99.7 | 100.0 | **2015** | 100.0 | 100.0 |
| **2004** | 97.8 | 99.2 | **2016** | 92.8 | 92.6 |
| **2005** | 90.9 | 96.4 | **2017** | 72.0 | 73.7 |
| **2006** | 76.5 | 82.2 | **2018** | 39.2 | 43.0 |
| **2007** | 55.7 | 55.3 | **2019** | 26.8 | 24.9 |

To level modern to Müller's observations, we discarded groups recorded at the Kislovodsk Observatory whose observed areas (not corrected by foreshortening) were between 10 and 100 millionths of the solar disc (msd). Because of possible limitations (for example, in the quality of the earliest telescopes), these constraints were applied so that observations made in the most recent solar cycles were levelled to those of Müller in 1709. We selected the aforementioned two values of sunspot areas because 100 and 10 msd represent typical thresholds for observer of low and high acuity, respectively (Usoskin et al. 2016). We did not apply these restrictions to SILSO data because that dataset does not include sunspot areas. Figure 2 (bottom panel) represents the ADF calculations from raw Kislovodsk data (blue dashed line) and those discarding groups whose observed areas were less than 10 and 100 msd (grey bars). In addition, grey points depict the computed ADF, discarding groups recorded in the Kislovodsk catalogue whose areas were lower than 50 msd. This provided us with an idea of an intermediate observer, that is, between one observer able to see groups with areas greater than 10 msd and one who could see only groups with observed areas greater than 100 msd. The most probable value of the ADF for 1709 and its upper and lower limits are also depicted as previously. Regarding the less conservative scenario (10 msd), the solar activity level that occurred in 2009 (24.8%) would be similar to the most probable value of the ADF for 1709 (25.1%), whereas the solar activity levels in 2008 (23.7%) and 2019 (22.3%) would be slightly lower. In addition, 2018 (30.4%) would have the highest level of solar activity as compared to the most probable value for 1709. When the most conservative constraint (100 msd) is applied, the solar activity level in 2009 (10.4%) and 2019 (11.3%) would be similar to the lower limit of the ADF for 1709 (11.3%), whereas that in 2018 (8.3%) would be slightly lower, and that in 2008 (5.3%) would be significantly lower. Applying an intermediate constraint of 50 msd to observed areas, the solar activity levels in 2008, 2009, 2018 and 2019 would be similar to the lower limit computed for the ADF in 1709.

We have converted the ADF for 1709 obtained in this work into values of sunspot number index. We have made this conversion by applying the method proposed by Kovaltsov et al. (2004): $SN = 19\ ADF^{1.25}$, where SN is the calculated value of the sunspot number and ADF the active day fraction obtained for 1709. This relation works well for active day fractions lower than 50 %. Thus, the value of the sunspot number in the case of the most probable of the ADF for 1709 (25.1%) is 3.4 whereas it is 5.8 in the case of the upper limit (38.9%). Applying the correction factor 1.67 generally used to scale old observations to the international sunspot number (version 2) to previous calculations, we obtain values of 5.6 and 9.7 according to the most probable value and the upper limit, respectively. The sunspot number value computed from the most probable value agrees with that found by Vaquero et al. (2015) for 1709 (up to around five in the strict model) and is significantly lower than that from the international sunspot number index (version 2), which is 13.3. Regarding the value for the upper limit, it is larger than that obtained by Vaquero et al. (2015) and lower than that from the international sunspot number (version 2).

Low solar activity during the Maunder Minimum

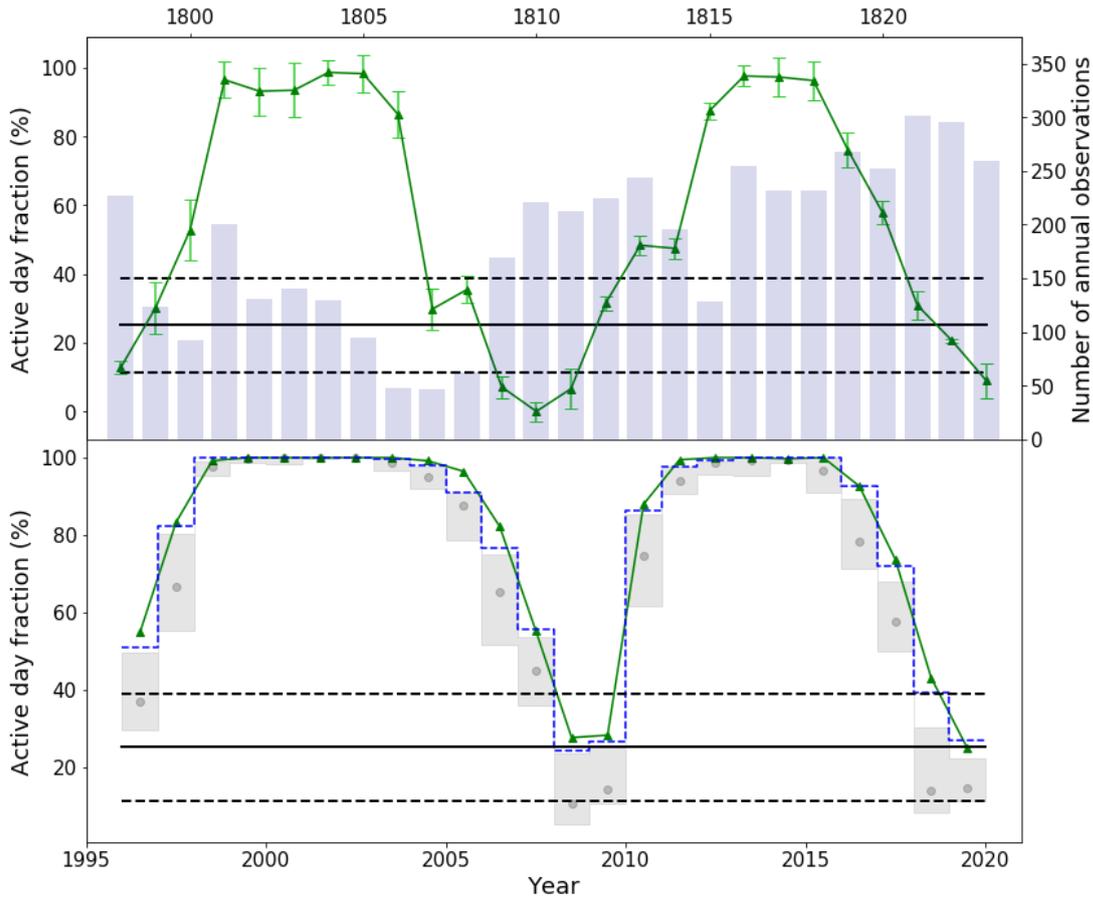

**Figure 2.** (Top panel) ADF calculated according to Vaquero et al. (2016) that includes Derfflinger's revised data presented in Hayakawa et al. (2020b) (green triangle-marked line) for the period 1798–1823. Error bars were calculated from hypergeometrical probability distribution. Dark blue bars indicate the number of daily observations per year. (Bottom panel) ADF calculated according to SILSO (green triangle-marked line) and raw data recorded at the Kislovodsk Observatory (dashed blue line) for the period 1996–2019. Grey bars represent the annual ADF range computed from Kislovodsk data, discarding groups whose observed areas were between 10 and 100 msd. Grey points depict the computed ADF, discarding groups whose observed areas were lower than 50 msd. Black horizontal lines in the top and bottom panels represent the most probable ADF value obtained from Müller's sunspot records from 1709, whereas horizontal black dashed lines indicate the upper and lower limits with a 99% significance level.

## 5. Conclusion

Some studies have shown problematic spotless days included in the sunspot databases for the MM. We have located a list produced by J. H. Müller that recorded spotless days in the second half of the year 1709. As Müller also recorded active days in that same year, this dataset provided us with a unique opportunity to determine the levels of solar activity with great precision, an otherwise extremely challenging task for other periods included in the MM. We therefore computed the most probable value and the upper and lower limits of the ADF for the year 1709 using the hypergeometrical probability distribution at a 99% significance level. We note that our sample is not strictly random because Müller observed in consecutive days when he recorded sunspots. Hence, our result depicts an upper threshold of solar activity for 1709.

We compared our results with the solar activity levels in other periods which showed a significant decrease in sunspot activity. In relation to the Dalton Minimum, 1709 was less active than most of the years of this period according to data from Vaquero et al. (2016) (after correction of Hayakawa et al. 2020b). Specifically, the most ADF probable value and the upper and lower limits obtained for 1709 would be in the 20th, 40th and 10th percentile in a ranking of the most active years of the Dalton Minimum. The years of the Dalton Minimum less active than 1709 correspond to years around the minima. Regarding the most recent solar cycles, we computed the ADF using data from the Kislovodsk sunspot catalogue and SILSO. According to raw data, 1709 was considerably less active than most of the recent years: no year was less active than the lower limit. Only the solar activity levels in 2008, 2009 and 2019 were similar to the most probable value for 1709. We also applied some constraints to the Kislovodsk data to ensure these records were on par with Müller's sunspot records due to possible limitations in Müller's observational capabilities (e.g., in detecting smaller sunspots). Thus, we applied two levels of constraints: one slightly conservative, which discarded groups whose observed areas were less than 10 msd, and another more conservative for 100 msd. In addition, we considered an intermediate constraint whereby groups were discarded whose observed areas were lower than 50 msd. Regarding the less conservative scenario, the solar activity level in 2009 was similar to that in 1709 according to its most probable value, and 2008 and 2019 were slightly less active. For the most conservative case, the solar activity levels in 2009 and 2019 were similar to the lower limits computed for 1709 and, moreover, 2018 and more significantly 2008 were even lower. In the case of the intermediate scenario, the solar activity levels that occurred in minima with the most solar cycles in 2008, 2009, 2018 and 2019 were similar to the lower limit corresponding to the ADF for 1709.

According to Vaquero et al. (2016), 1709 was a rather active year in the MM. Indeed, it was the ninth year with the highest number of active days and 13th most active year according to the ADF. In fact, 1709 was not included in the core period of the MM but in the transition phase to 'normal activity' according to Vaquero & Trigo (2015). We concluded that although 1709 was one of the most active years in the MM, it was still less active than most years in the Dalton Minimum and those of the most recent solar cycles. In fact, it was comparable to the solar activity levels of years around minima in the Dalton Minimum and at least similar to the current solar minima of 2008–2009 and 2018–2019. This is consistent with the apparent loss of the solar coronal streamers during the MM, in contrast with the visible streamers in the Dalton Minimum and latest solar cycles (Hayakawa et al. 2020a,b). These facts constitute strong evidence for low solar activity levels in the MM despite the latest published works that cast doubt on this observational fact (Zolotova & Ponyavin 2015). Furthermore, we converted the ADF values into sunspot number values. Then, applying the factor 1.67 to scale our results to the international sunspot number (version 2), the most probable value obtained in this work agrees with values of the sunspot number obtained by Vaquero et al. (2015) and is significantly lower than the value of the international sunspot number (version 2). In the case of the value according to the upper limit, it is larger than that obtained by Vaquero et al. (2015) and lower than that from the international sunspot number (version 2). A broader study will be carried out in future works to assess a possible overestimation in the international sunspot number index (version 2) for the solar cycle around 1700–1710.

**Data Availability Statement**

Data used in this work on the number of groups recorded by any observer during the Maunder Minimum is available through Vaquero et al. (2016), the number of quiet and active days from Müller's observations is available through Hayakawa et al. (2021d), the number of quiet and active days from SILSO is available at http://www.sidc.be/silso/ and data from Kislovodsk Observatory at http://158.250.29.123:8000/web/Soln_Dann/.

**Acknowledgements**

This research was supported by the Junta of Extremadura through grant GR18097 (co-financed by the European Regional Development Fund) and by the Ministerio de Economía y Competitividad of the Spanish Government (CGL2017-87917-P). HH and CK's studies have been conducted in the frameworks of JSPS Grant-in-Aids JP15H05812, JP17J06954, JP20K22367, JP20K20918, and JP20H05643, JSPS Overseas Challenge Program for Young Researchers, the 2020 YLC collaborating research fund, and the research grants for Mission Research on Sustainable Humanosphere from Research Institute for Sustainable Humanosphere (RISH) of Kyoto University and Young Leader Cultivation (YLC) program of Nagoya University. The authors thank the National Library of Russia for letting them consult J.H. Müller's original manuscripts, SILSO for providing international sunspot number, and Dr. Andrey Tlatov for his useful help by providing an update of the Kislovodsk catalogue. Finally, the authors have benefited from the participation in the ISSI workshops led by M.J. Owens and F. Clette on the calibration of the sunspot number.

**Disclosure of Potential Conflicts of Interest and Ethical Statement**

The authors declare that they have no conflicts of interest.

**Author Contributions**

All authors contributed to the study conception and design. The first draft of the manuscript was written by V.M.S. Carrasco, the original historical records have been consulted and analysed by H. Hayakawa and C. Kuroyanagi, and all authors read and approved the final manuscript.

V.M.S. Carrasco et al.